\documentclass[runningheads,a4paper]{llncs}
\usepackage{amsmath,latexsym}
\usepackage{amsfonts}
\usepackage{amssymb}
\usepackage{graphicx}

\usepackage{subfigure}
\usepackage{color}
\usepackage{epstopdf} 
\usepackage{setspace}
\usepackage{algpseudocode}
\usepackage{algorithm}
\usepackage{cite}

\usepackage{tikz}
\usetikzlibrary{automata,positioning}
\usetikzlibrary{arrows,automata}
\usetikzlibrary{arrows,calc}
\usetikzlibrary{through}
\usetikzlibrary{decorations,decorations.markings,decorations.text}
\usetikzlibrary{decorations.pathmorphing}
\usetikzlibrary{patterns}

\def\qedbox#1#2{\vbox{\hrule height.2pt
  \hbox{\vrule width.2pt height#2pt \kern#1pt \vrule width.2pt}
  \hrule height.2pt}}

\def\s#1{\mbox{\boldmath $#1$}}

\def\+{\!+\!}
\def\-{\!-\!}

\def\m{\!-\!}

\def\itbf#1{\textit{\textbf{#1}}}
\def\match{\approx}

\def\bproc{{\bf procedure\ }}

\def\bfor{{\bf for\ }}
\def\bto{{\bf to\ }}
\def\bdownto{{\bf downto\ }}
\def\bwhile{{\bf while\ }}

\def\band{{\bf and\ }}
\def\bor{{\bf or\ }}
\def\bdo{{\bf do\ }}
\def\bif{{\bf if\ }}
\def\bthen{{\bf then\ }}
\def\belse{{\bf else\ }}

\def\bbreak{{\bf break\ }}
\def\la{\leftarrow}

\def\qq{\qquad}
\def\com#1{\hspace{24pt}{\bf $\triangleright$}\hspace{6pt}{\sl #1}}

\def\pref(#1,#2){$#1$ is a prefix of $#2$}
\def\suff(#1,#2){$#1$ is a suffix of $#2$}

\def\reg(#1,#2){$#2$ is $#1$-regular}
\def\notreg(#1,#2){$#2$ is not $#1$-regular}
\def\top{\tt{top}}
\def\pop{\tt{pop}}
\def\push{{\tt{push}}}

\def\UPDATE\_F{\tt{UPDATE\_F}}

\def\PCInd{\tt{PCInd}}
\def\PCR{\tt{PCR}}
\newcommand{\dd}{\mathinner{\ldotp\ldotp}}

\algnewcommand{\LineComment}[1]{\State \(\triangleright\) \normalfont{\sl #1}}
\algtext*{EndWhile}% l
\algtext*{EndIf}% Remove "end if" text
\algtext*{EndFor}% Remove "end for" text
\algtext*{EndFor}% Remove "end for" text
\algtext*{EndProcedure}% Remove "end procedure" text

\begin{document}
\pagestyle{headings}
\title{Computing Covers Using Prefix Tables
\thanks{This work was supported in part by the Natural Sciences \& Engineering
Research Council of Canada.}
}

\titlerunning{\itshape{Computing Covers Using Prefix Tables}}

\author{
Ali Alatabbi\inst{1}
\and 
M.\ Sohel Rahman
\thanks{Partially supported by a Commonwealth Academic Fellowship and
an ACU Titular Fellowship, both funded by the UK Government. Currently
on a sabbatical leave from BUET.}\inst{2}
\and 
W.\ F.\ Smyth\inst{1,3,4}
}

\authorrunning{\itshape{Alatabbi, Rahman and Smyth.}}

\institute{Department of Informatics, King's College London\\
\email{ali.alatabbi@kcl.ac.uk} 
\and Department of Computer Science \& Engineering\\
Bangladesh University of Engineering \& Science\\
\email{msrahman@cse.buet.ac.bd}
\and Algorithms Research Group, Department of Computing \& Software\\
McMaster University\\
\email{smyth@mcmaster.ca} 
\and School of Engineering \& Information Technology\\
Murdoch University, Western Australia}

\maketitle
\begin{abstract}
An \itbf{indeterminate string}
$\s{x} = \s{x}[1..n]$ on an alphabet $\Sigma$
is a sequence of nonempty subsets of $\Sigma$;
\s{x} is said to be \itbf{regular} if every subset is of size one.
A proper substring \s{u} of regular \s{x} is said to be a \itbf{cover} of \s{x}
iff for every $i \in 1..n$,
an occurrence of \s{u} in \s{x} includes $\s{x}[i]$.
The \itbf{cover array} $\s{\gamma} = \s{\gamma}[1..n]$ of \s{x}
is an integer array such that $\s{\gamma}[i]$ is the longest cover
of $\s{x}[1..i]$.
Fifteen years ago a complex, though nevertheless linear-time, algorithm
was proposed to compute the cover array of regular \s{x}
based on prior computation of the border array of \s{x}.
In this paper we first describe a linear-time algorithm
to compute the cover array of regular \s{x} based on the prefix table of \s{x}.
We then extend this result to indeterminate strings.
% as well as to the computation of left seeds of regular and indeterminate strings.
\end{abstract}

\section{Introduction}\label{sect-intro}
The idea of a \itbf{quasiperiod} or \itbf{cover} of a string \s{x}
was introduced almost a quarter-century ago by
Apostolico \& Ehrenfeucht \cite{AE90}:
a proper substring \s{u} of \s{x} such that every position in \s{x}
lies within an occurrence of \s{u}.
Thus, for example, $\s{u} = aba$ is a cover of
$\s{x} = ababaababa$.
In \cite{AFI91} a linear-time algorithm was described
to compute the shortest cover of \s{x};
this contribution was followed by linear-time algorithms to compute

\begin{itemize}
\item[$\bullet$]
the shortest cover of every prefix of \s{x} \cite{B92};
\item[$\bullet$]
all the covers of \s{x} \cite{MS94,MS95};
\item[$\bullet$]
all the covers of every prefix of \s{x} \cite{LS02}.
\end{itemize}

A \itbf{border} of a string \s{x} is a possibly empty
proper prefix of \s{x} that is also a suffix of \s{x}.
(Thus a cover of \s{x} is necessarily also a border of \s{x}.)
%A string \s{x} of length $n$ has a border of length $b$
%if and only if it has \itbf{period} $n\- b$.
In the \itbf{border array} $\s{\beta} = \s{\beta}[1..n]$
of the string $\s{x} = \s{x}[1..n]$,
$\s{\beta}[i]$ is the length of the longest border of $\s{x}[1..i]$.
Since for $\s{\beta}[i] \ne 0$, $\s{\beta}[\s{\beta}[i]]$
is the length of a border of \s{x} as well as
the length of the longest border of $\s{x}[1..\s{\beta}[i]]$
\cite{AHU74,S03},
it follows that $\s{\beta}$ provides all the borders
of every prefix of \s{x}.
For example:

\begin{equation}
\begin{array}{rcccccccccc}
\scriptstyle 1 & \scriptstyle 2 & \scriptstyle 3 & \scriptstyle 4 & \scriptstyle 5 & \scriptstyle 6 & \scriptstyle 7 & \scriptstyle 8 & \scriptstyle 9 & \scriptstyle 10 \\
\s{x} = a & b & a & b & a & b & a & a & b & a \label{x} \\
\s{\beta} = 0 & 0 & 1 & 2 & 3 & 4 & 5 & 1 & 2 & 3
\end{array}
\end{equation}

As shown in \cite{LS02}, the \itbf{cover array} \s{\gamma}
has a similar cascading property, giving the lengths
of all the covers of every prefix of \s{x} in a compact form:

\begin{equation*}
\begin{array}{rcccccccccc}
\scriptstyle 1 & \scriptstyle 2 & \scriptstyle 3 & \scriptstyle 4 & \scriptstyle 5 & \scriptstyle 6 & \scriptstyle 7 & \scriptstyle 8 & \scriptstyle 9 & \scriptstyle 10 \\
\s{\gamma} = 0 & 0 & 0 & 2 & 3 & 4 & 5 & 0 & 0 & 3
\end{array} 
\end{equation*} 

Here $\s{x}[1..7]$ has covers $\s{u_1} = \s{x}[1..5] = ababa$
and $\s{u_2} = \s{x}[1..3] = aba$,
while the entire string \s{x} has cover \s{u_2}.
The main result of \cite{LS02} is an algorithm that computes
$\s{\gamma} = \s{\gamma}[1..n]$ from $\s{\beta} = \s{\beta}[1..n]$
in $\Theta(n)$ time, while making no reference
to the underlying string \s{x}.

The results outlined above all apply to a \itbf{regular string} --- that
is, a string \s{x} such that each entry $\s{x}[i]$ is constrained to be a one-element
subset of a given set $\Sigma$ called the \itbf{alphabet}.
In this paper we show how to extend these ideas and algorithms
to an \itbf{indeterminate string} \s{x} --- that is,
such that each $\s{x}[i]$ can be any nonempty subset of $\Sigma$.
Observe that every regular string is indeterminate.

The idea of an indeterminate string was first introduced in
\cite{FP74}, then studied further in the 1980s as a ``generalized string''
\cite{A87}.
Over the last 15 years Blanchet-Sadri has written
numerous papers on the properties of ``strings with holes''
(each $\s{x}[i]$ is either a one-element subset of $\Sigma$
or $\Sigma$ itself),
together with a monograph on the subject \cite{BS07};
while other authors have studied indeterminate strings
in their full generality, together with related algorithms
\cite{BRS09,NRR12,HS03,HSW08,SW08,SW09a,SW09,CRSW13}.
In the specific context of this paper, Vor\'{a}\v{c}ek \& Melichar \cite{VM05}
have done pioneering work on the computation of covers and related structures
in generalized strings using finite automata.

For indeterminate strings, equality of letters
is replaced by the idea of a ``match'' \cite{HS03}:
$\s{x}[i]$ \itbf{matches} $\s{x}[j]$
(written $\s{x}[i] \approx \s{x}[j]$) if and only if
$\s{x}[i] \cap \s{x}[j] \ne \emptyset$,
while $\s{x} \approx \s{y}$ if and only if
$|\s{x}| = |\s{y}|$ and corresponding positions in
\s{x} and \s{y} all match.
It is important to note that matching is nontransitive:
$b \approx \{b,c\} \approx c$, but $b \not\approx c$.

It is \cite{CRSW13} that provides the point of departure for our contribution,
as we now explain.
The \itbf{prefix table} $\s{\pi} = \s{\pi}[1..n]$ of $\s{x}[1..n]$ is an
integer array such that $\s{\pi}[1] = n$ and,
for every $i \in 2..n$, $\s{\pi}[i]$ is the length of the longest substring
occurring at position $i$ of \s{x} that matches a prefix of \s{x}.
Thus, for our example (\ref{x}):

\begin{equation*}
\begin{array}{rccccccccccc}
%BILL
& \scriptstyle 1 & \scriptstyle 2 & \scriptstyle 3 & \scriptstyle 4 & \scriptstyle 5 & \scriptstyle 6 & \scriptstyle 7 & \scriptstyle 8 & \scriptstyle 9 & \scriptstyle 10 \\
\s{x} 	= & a & b & a & b & a & b & a & a & b & a\\
\s{\pi} = & 10 & 0 & 3 & 0 & 3 & 0 & 1 & 3 & 0 & 1
\end{array}
\end{equation*} 

It turns out \cite{BKS13} that the prefix table and the border array
are ``equivalent'' for regular strings; that is,
each can be computed from \s{x} in linear time,
and each can be computed from the other,
without reference to \s{x}, also in linear time.
However, for indeterminate strings, this is not true:
the prefix table
continues to determine all the borders of every prefix of \s{x},
while the border array, due to the intransitivity of matching,
is no longer reliable in identifying borders shorter than the longest one.
Consider, for example:

\begin{equation*}
\begin{array}{rccc}
\scriptstyle 1 & \scriptstyle 2 & \scriptstyle 3 \\
\s{x} = a & \{a,b\} & b \\
\s{\beta} = 0 & 1 & 2
\end{array}
\end{equation*}

Here \s{x} does not have a border of length $\s{\beta}[\s{\beta}[3]] = 1$;
on the other hand, $\s{\pi} = 320$ correctly identifies
all the borders of every prefix of \s{x}.

%\add[S]{This is why in the literature the border array of an indeterminate string has been given an extended definition}\cite{HS03,NRR12,BRS09}\add[S]{: unlike the border array of a regular string, which is a simple array of integers, the same for indeterminate string is an array of lists of integers. Here at each position, the list gives all possible borders for that prefix.}

Moreover, it was shown in \cite{CRSW13} that every \itbf{feasible} array --- that is,
every array $\s{y} = \s{y}[1..n]$ such that $\s{y}[1] = n$
and for every $i \in 2..n$, $\s{y}[i] \in 0..n\- i\+ 1$ --- is a
prefix table of some (indeterminate) string.
Thus there exists a many-many correspondence between all possible
prefix tables and all possible indeterminate strings.
Furthermore, \cite{SW08} describes an algorithm to compute
the prefix table of any indeterminate string,
while \cite{ARS14} gives an algorithm to compute a
lexicographically least indeterminate string
corresponding to a given prefix table.

At this point let us discuss our motivation more precisely. First,
realize that to exploit the fullest functionality of a border array of an
indeterminate string we need to resort to the extended definition of the border
array which in fact requires quadratic space \cite{HS03,NRR12,BRS09}:
unlike the border array of a regular string, which is a simple array of
integers, the border array of an indeterminate string is an array of lists of
integers. Here at each position, the list gives all possible borders for that
prefix. On the other hand, the prefix array, even for the indeterminate string,
remains a simple one-dimensional array, just as for a regular string. It thus
becomes of interest to make use of the prefix table rather than the border array
whenever possible, in order to extend the scope of computations to indeterminate
strings.

In Section~\ref{sect-pcr} of this paper, we describe a linear-time
algorithm to compute the cover array \s{\gamma}
of a regular string \s{x} directly from its prefix table \s{\pi}. Then,
Section~\ref{sect-indet} describes a limited extension of this algorithm to
indeterminate strings.
Finally, Section~\ref{sect-future} outlines future research directions,
especially making use of prefix tables to extend the utility and applicability
of other data structures to indeterminate strings.

\section{Prefix-to-Cover for a Regular String}\label{sect-pcr}
In this section we describe our basic $\Theta(n)$-time Algorithm $\PCR$ to
compute the cover array $\s{\gamma} = \s{\gamma}[1..n]$ of a regular string
$\s{x} = \s{x}[1..n]$ directly from its prefix table $\s{\pi} = \s{\pi}[1..n]$.
In fact, as noted in the Introduction, \s{\gamma} actually provides
all the covers of every prefix of \s{x}.
Central to our algorithm are the following definitions:

\begin{definition}\label{defn-range}
If, for a position $i \in 1..n$, $\s{\pi}[i] > 0$,
then $R_i = [i,i\+ \s{\pi}[i]\- 1]$ is said to be
the \itbf{range} at $i$ of \itbf{length} $\s{\pi}[i]$;
the ranges $R_i$ and $R_{i'}$, $i' > i$, are \itbf{connected}
if and only if $i' \le i\+ \s{\pi}[i] < i'\+ \s{\pi}[i']$.
\end{definition}
 
Notably, in what follows, for the sake of brevity, we may slightly abuse the
notation $R_i = [i,i\+ \s{\pi}[i]\- 1]$ by simply saying $R_i = \s{\pi}[i]$.

\begin{definition}\label{defn-live}
Position $j$ in \s{\pi} is said to be \itbf{live} at position $i' > j$ if and only if
there exists a sequence of $h \ge 1$ connected ranges
$R_{i_1}, R_{i_2}, \ldots, R_{i_h}$, each of length at least $j$, such that
$i_1 \le j\+ 1,\ i_h\+ \s{\pi}[i_h]\- 1\ge i'$.
Otherwise, $j$ is said to be \itbf{dead} at $i'$.
\end{definition}

Thus $\s{x}[1..n]$ has a cover $\s{x}[1..j]$, $j < n$, if and only if
$j$ is live at $n$ and the final connected range $R_{i_h}$
satisfies $i_h\+ \s{\pi}[i_h]\- 1 = n$.

The strategy of Algorithm $\PCR$ (Figure~\ref{alg-pcr})
is to perform an on-line left-to-right scan of \s{\pi},
identifying connected ranges $R_i$.
This process may be complex.
Within range $R_i$ there may exist two (or more) positions $i_1 >i$ and $i_2 > i_1$
that define ranges $R_{i_1}$ and $R_{i_2}$, both connected to $R_i$;
of these, $\PCR$ processes $R_i$ first, followed by $R_{i_1}$,
then, if $R_{i_1}$ and $R_{i_2}$ are connected (they may not be),
by $R_{i_2}$.
For example, consider\footnote{Thanks to Alice Heliou,
Laboratoire d'Informatique de l'\'{E}cole Polytechnique, Palaiseau, France.}

\begin{equation}
\begin{array}{r cccccccccccccccccccc}
& \scriptstyle 1 & \scriptstyle 2 & \scriptstyle 3 & \scriptstyle 4 &
\scriptstyle 5 & \scriptstyle 6 & \scriptstyle 7 & \scriptstyle 8 & \scriptstyle 9 & \scriptstyle 10 & \scriptstyle 11 & \scriptstyle 12 & \scriptstyle 13 & \scriptstyle 14 & \scriptstyle 15 & \scriptstyle 16 & \scriptstyle 17 & \scriptstyle 18 & \scriptstyle 19 \\
\s{x} = & b & a & b & a & b & a & b & b & a & b & a & b & a & b & a & b & a
& b & a \label{alice} \\
\s{\pi} = & 19 & 0 & 5 & 0 & 3 & 0 & 1 & 7 & 0 & 7 & 0 & 7 & 0 & 6 & 0 & 4
& 0 & 2 & 0 \\ 
\s{\gamma} = & 0 & 0 & 0 & 2 & 3 & 4 & 5 & 0 & 0 & 3 & 0 & 5 & 0 & 7 & 0 &
7 & 0 & 7 & 0
\end{array}
\end{equation}

Here the pairs of ranges $(R_8,R_{10})$, $(R_8,R_{12})$ and $(R_{10},R_{12})$
are all connected:
$\PCR$ will process positions 8--14 in $R_8$,
followed by 15--16 in $R_{10}$, then 17--18 in $R_{12}$
and finally position 19 in $R_{14}$.

\begin{figure}[htbp]
{\leftskip=0.15in\obeylines\sfcode`;=3000
\bproc $\PCR$ $(\s{\pi},\s{\gamma})$
$\s{\gamma}[1..n] \la 0^n;\ maxlive[1..n] \la 0^n$
$lastlim \la 1;\ i \la 2$
\bwhile $lastlim < n$ \bdo
\qq $j \la \s{\pi}[i]$
\qq \bif $j = 0$ \bthen
\com{No range extends beyond $lastlim$, so $1,2,\ldots,i\m 1$ are all dead.}
\qq\qq \bif $i > lastlim$ \bthen
\qq\qq\qq $maxlive[i\m 1] \la -1;\ lastlim \la i$
\qq \belse
\qq\qq $\lim \la i\+ j\m 1$
\qq\qq \bif $lim > lastlim$ \bthen
\qq\qq\qq $j' \la (lastlim\+ 1) - i$
\qq\com{Initial setting of $maxlive$ and \s{\gamma}.}
\qq\qq\qq \bfor $i' \la lastlim\+ 1$ \bto $lim$ \bdo
\qq\qq\qq\qq $j' \la j'\+ 1$
\qq\qq\qq\qq \bif ($maxlive[j'] = 0$ \band $i' \le 2j'$)
\qq\qq\qq\qq \bor $maxlive[j'] \ge i'\m j'$ \bthen
\qq\qq\com{$j'$ is a cover of $\s{x}[1..i']$.}
\qq\qq\qq\qq\qq $maxlive[j'] \la i';\ \s{\gamma}[i'] \la j'$
\qq\qq\qq\qq \belse
\qq\qq\com{$j'$ is ruled out as a cover.}
\qq\qq\qq\qq\qq $maxlive[j'] \la -1$
\qq\com{Reset $maxlive$ and \s{\gamma} in case of multiple covers.}
\qq\qq\qq \bfor $i' \la lim$ \bdownto $lastlim\+ 1$ \bdo
\qq\qq\qq\qq $j'' \la \s{\gamma}[j']$
\qq\qq\com{A cover of $\s{x}[1..j']$ is also a cover of $\s{x}[1..i']$.}
\qq\qq\qq\qq \bwhile $j'' > 0$ \band $0 < maxlive[j''] < i'$ \bdo
\qq\qq\qq\qq\qq $maxlive[j''] \la i';\ \s{\gamma}[i'] \la \max(\s{\gamma}[i'],j'')$
\qq\qq\qq\qq\qq $j'' \la \s{\gamma}[j'']$
\qq\qq\qq\qq $j' \la j'\m 1$
% \qq\qq\qq\qq \belsif $maxlive[\j'] \ge i'\m 1$ \bthen
% \qq\qq\qq\com{$\s{x}[1..i']$ is covered by $\s{x}[1..\j']$.}
% \qq\qq\qq\qq\qq $maxlive[\j'] \la i';\ \s{\gamma}[i'] \la \j'$
% \qq\qq\qq\qq \belsif $maxlive[\j'] > 0$ \bthen
% \qq\qq\qq\qq\qq $\j'' \la \s{\gamma}[\j']$
% \qq\qq\qq\com{$\j'$ becomes dead: check its ancestor $\j''$.}
% \qq\qq\qq\qq\qq SETANCESTORS $(maxlive,\s{\gamma},i',\j',\j'')$
% \qq\qq\qq\qq \belsif $maxlive[\j'] < -1$ \bthen
% \qq\qq\qq\qq\qq $\j'' \la -maxlive[\j']$
% \qq\qq\qq\com{$\j'$ already dead; check stored ancestor $\j''$.}
% \qq\qq\qq\qq\qq SETANCESTORS $(maxlive,\s{\gamma},i',\j',\j'')$
\qq\qq\qq $lastlim \la lim$
\qq $i \la i\+ 1$
}
\caption{Compute the cover array $\s{\gamma}$ of a regular string \s{x} from its prefix table $\s{\pi}$.}
\label{alg-pcr}
\end{figure}

Algorithm $\PCR$ processes each connected range $R_i$ twice,
first in left-to-right order,
beginning at position $i' = lastlim\+ 1$,
where $lastlim$ is the current rightmost position for which $\s{\gamma}$
has already been determined,
and ending at $i' = lim > lastlim$, the rightmost position in $R_i$.
Corresponding to each $i'$ is the length $j' = i'\- i\+ 1$ of the
prefix of $R_i$ (hence also of \s{x}) that may extend a sequence of
covering substrings of length $j'$.
In order to determine whether or not $j'$ is live at $i'$,
$\PCR$ maintains an array $maxlive[1..n]$,
using the following values:

\begin{eqnarray*}
maxlive[j'] = 0& : & \mbox{initial setting: position $j'$ not yet considered} \\
i'& : & \mbox{$j'$ live at $i'$: $\s{x}[1..i']$ covered by $\s{x}[1..j']$} \\
-1& : & \mbox{$j'$ is (permanently) dead}
\end{eqnarray*}

However, it can happen that $maxlive$ and \s{\gamma} are not correctly set
by the left-to-right scan of $R_i$:

\begin{definition}[\cite{LS02}]
\label{defn-ancestor}
In the cover array \s{\gamma},
if there exists an integer $k \ge 1$ and positions $i > j > 0$
such that $\s{\gamma}^k[i] = j$,
then $j$ is said to be the \itbf{$k^{\mbox{th}}$ ancestor} of $i$ in \s{\gamma}.
Thus the cover array determines a \itbf{cover tree}.
\end{definition}

It may be that $\s{\gamma}[i']$ is set to zero because $j'$ is dead at $i'$,
even though an ancestor of $j'$ in the cover tree is live at $i'$;
on the other hand, when $\s{\gamma}[i'] = j'$,
so that ancestors of $j'$ may also be live at $i'$,
the $maxlive$ values of the ancestors may need to be adjusted.
Thus a second right-to-left scan of $R_i$ is required,
in order to ensure that these updates are correct.

For example, in (\ref{alice}),
we need to ensure that $maxlive[5] = maxlive[3] = 18$,
since both 5 and 3 are live ancestors of 7.
A more subtle example is given in (\ref{alice'}),
where at position 19 we need to recognize that both 5 and 3 are live,
even though 7 is dead,
so that later, at position 22, we can recognize that 3 is live:

\begin{equation}
\begin{array}{rccccccccccccccccccccccc}
& \scriptstyle 1 & \scriptstyle 2 & \scriptstyle 3 & \scriptstyle 4 &
\scriptstyle 5 & \scriptstyle 6 & \scriptstyle 7 & \scriptstyle 8 & \scriptstyle 9 & \scriptstyle 10 & \scriptstyle 11 & \scriptstyle 12 & \scriptstyle 13 & \scriptstyle 14 & \scriptstyle 15 & \scriptstyle 16 & \scriptstyle 17 & \scriptstyle 18 & \scriptstyle 19 & \scriptstyle 20 & \scriptstyle 21 & \scriptstyle 22 \\
\s{x} = & b & a & b & a & b & a & b & b & a & b & a & b & b & a & b & a & b
& a & b & b & a & b \label{alice'} \\
\s{\pi} = & 22 & 0 & 5 & 0 & 3 & 0 & 1 & 5 & 0 & 3 & 0 & 1 & 7 & 0 & 5 & 0
& 3 & 0 & 1 & 3 & 0 & 1 \\
\s{\gamma} = & 0 & 0 & 0 & 2 & 3 & 4 & 5 & 0 & 0 & 3 & 0 & 5 & 0 & 0 & 3 &
0 & 5 & 0 & 5 & 0 & 0 & 3\\
\end{array}
\end{equation} 

Consider also

\begin{equation}
\begin{array}{rccccccccccccccccccccccccccccc}
& \scriptstyle 1 & \scriptstyle 2 & \scriptstyle 3 & \scriptstyle 4 &
\scriptstyle 5 & \scriptstyle 6 & \scriptstyle 7 & \scriptstyle 8 & \scriptstyle 9 & \scriptstyle 10 & \scriptstyle 11 & \scriptstyle 12 & \scriptstyle 13 & \scriptstyle 14 & \scriptstyle 15 & \scriptstyle 16 & \scriptstyle 17 & \scriptstyle 18 & \scriptstyle 19 & \scriptstyle 20 & \scriptstyle 21 & \scriptstyle 22 & \scriptstyle 23 & \scriptstyle 24 & \scriptstyle 25 & \scriptstyle 26 & \scriptstyle 27 & \scriptstyle 28 \\
\s{x} = & b & a & b & a & b & a & b & b & a & b & a & b & a & b & a & b & b
& a & b & a & b & b & a & b & a & b & a & b \label{alice''} \\
\s{\pi} = & 22 & 0 & 5 & 0 & 3 & 0 & 1 & 7 & 0 & 7 & 0 & 5 & 0 & 3 & 0 & 1
& 5 & 0 & 3 & 0 & 1 & 7 & 0 & 5 & 0 & 3 & 0 & 1 \\
\s{\gamma} = & 0 & 0 & 0 & 2 & 3 & 4 & 5 & 0 & 0 & 3 & 0 & 5 & 0 & 7 & 0 &
7 & 0 & 0 & 3 & 0 & 5 & 0 & 0 & 3 & 0 & 5 & 0 & 5\\
\end{array}
\end{equation}

Thus, using $n$ additional words of storage and a double scan of each connected range,
Algorithm $\PCR$ is able to compute $\s{\gamma}$.
The time requirement is $\Theta(2n)$ plus the time required by the internal \bwhile loop;
this loop updates $maxlive[j']$ at most once for each ancestral position $j'$ in the range,
thus requiring a total $O(n)$ time overall. Hence we have the following result:

\bigskip
\begin{theorem}
Given the prefix table $\s{\pi}$ of a regular string $\s{x} = \s{x}[1..n]$,
Algorithm $\PCR$ correctly computes the cover array $\s{\gamma}$ of \s{x}
in $\Theta(n)$ time using an additional $n$ integers of space.
\end{theorem}

\begin{equation}
\label{org-string}
\begin{array}{ r c c c c c c c c c c c c c c c c c c c c c c c}
 &\scriptstyle1 & \scriptstyle 2 & \scriptstyle 3 & \scriptstyle
4 & \scriptstyle 5 & \scriptstyle 6 & \scriptstyle 7 & \scriptstyle
8 & \scriptstyle 9 & \scriptstyle 10 & \scriptstyle 11 & \scriptstyle
12 & \scriptstyle 13 & \scriptstyle 14 & \scriptstyle 15 & \scriptstyle
16 & \scriptstyle 17 & \scriptstyle 18 & \scriptstyle 19 & \scriptstyle
20 & \scriptstyle 21 & \scriptstyle 22 & \scriptstyle 23 \\
%\hline
\s{x} = &a&b&a&a&b&a&b&a&a&b&a&a&b&a&b&a&a&b&a&b&a&b&a\\
\pi  = &23&0&1&3&0&6&0&1&11&0&1&3&0&8&0&1&3&0&3&0&3&0&1\\
\gamma = &0&0&0&0&0&3&0&3&0&5&6&0&5&6&0&8&9&10&11&0&8&0&3\\
\end{array} 
\end{equation}

%\caption{The prefix and cover array of $\s{x}=abaababaabaababaabababa$}.
%\end{figure}

\begin{figure}[H]
\centering 
\begin{tikzpicture}[thick,scale=0.6]
\draw[style=thick] (0,0) -- (15,0);
\draw[style=thick] (0,1) -- (15,1);

\draw[decorate, decoration={snake, segment length=4mm, amplitude=.5mm}]
(0,0) -- (0,1);
\draw[decorate, decoration={snake, segment length=4mm, amplitude=.5mm}]
(15,0) -- (15,1);
\draw[style=thick] (3.5,0) rectangle (9.5,1)node[pos=0.5] {$\s{x}[1\dd c']$};

\draw[|<->|,style=thick] (3.5,.5) -- (5.5,.5) node[midway,fill=white] {$c$};
\draw[|<->|,style=thick] (7.5,.5) -- (9.5,.5) node[midway,fill=white] {$c$};

\draw[style=thick,pattern=north east lines,pattern color=gray!30]
(0.25,1) rectangle (2.25,1.75)node[pos=0.5] {$R_{i}$};

\draw[style=dashed] (.25,1.75) rectangle (2.25,2.5)node[pos=0.5] {$\s{x}[1\dd c]$};

\draw[style=thick,pattern=north east lines,pattern color=gray!30]
(1.5,0) rectangle (5.5,-.75)node[pos=0.5] {$R_{i'}$};

\draw[style=dashed] (1.5,-.75) rectangle (3.5,-1.5)node[pos=0.5] {$\s{x}[1\dd c]$};

\draw[style=thick,pattern=north east lines,pattern color=gray!30]
(3.5,1) rectangle (11,1.75)node[pos=0.5] {$R_{i''}$};

\draw[style=dashed] (3.5,1.75) rectangle (5.5,2.5)node[pos=0.5] {$\s{x}[1\dd c]$};
\draw[style=dashed] (7.5,1.75) rectangle (9.5,2.5)node[pos=0.5] {$\s{x}[1\dd c]$};

\draw[|<->|,style=thick] (3.5,3) -- (9.5,3) node[midway,fill=white] {$c'$};

\draw[style=thick,pattern=north east lines,pattern color=gray!30]
(7.5,0) rectangle (13.5,-.75)node[pos=0.5] {$R_{i'''}$};
\draw[style=dashed] (7.5,-.75) rectangle (9.5,-1.5)node[pos=0.5] {$\s{x}[1\dd c]$};
\draw[style=dashed] (11.5,-.75) rectangle (13.5,-1.5)node[pos=0.5] {$\s{x}[1\dd c]$};

\draw[|<->|,style=thick] (7.5,-2) -- (13.5,-2) node[midway,fill=white] {$c'$};

\draw[style=thick,pattern=north east lines,pattern color=gray!30]
(13,1) rectangle (15,1.75)node[pos=0.5] {$R_{i''''}$};

\draw[style=dashed] (13,1.75) rectangle (15,2.5)node[pos=0.5] {$\s{x}[1\dd c]$};

\end{tikzpicture} 
\label{pcr-view}
\caption{
Showing two covers from $\gamma(\s{x})$, 
$\s{x}=abaababaabaababaabababa$ (\ref{org-string})}
\end{figure}

Given a string $\s{x}=abaababaabaababaabababa$ (\ref{org-string}),
the figure shows two covers from $\gamma(\s{x})$, namely $c=3$ and $c'=8$,
%BILL
and also shows selected ranges from the prefix array $\pi(\s{x})$ that
%BILL
explicitly participate in the generation of these covers: the ranges are $
R_{i} = \pi[4] = 3,~ R_{i'}= \pi[6] = 6,~ R_{i''} = \pi[9]=11,~ R_{i'''} =
\pi[14]=8,~ R_{i''''} = \pi[21] = 3$.
\section{Extensions to Indeterminate Strings}
\label{sect-indet}
It turns out that for indeterminate strings there are
two natural analogues of the idea of ``cover''.

\begin{definition}
\label{defn-sliding}
A string $\s{x} = \s{x}[1..n]$ is said to have a \itbf{sliding cover}
of length $\kappa$
if and only if
\begin{itemize}
\item[(a)]
\s{x} has a suffix \s{v} of length $|\s{v}|= \kappa$; and
\item[(b)]
\s{x} has a proper prefix \s{u}, $|\s{u}| \ge |\s{x}|\- \kappa$,
with suffix $\s{v'} \match \s{v}$; and
\item[(c)]
either $\s{u} = \s{v'}$ or else \s{u} has a cover of length $\kappa$.
\end{itemize}
\end{definition}

A sliding cover requires that adjacent or overlapping substrings of \s{x}
match, but the nontransitivity of matching leaves open the possibility
that nonadjacent elements of the cover do not match.
For example,
\begin{equation}
\label{x'}
\s{x} = \{a,b\}c\{a,c\}\{a,c\}ca
\end{equation}
has a sliding cover of length $\kappa = 2$ because $\{a,b\}c \match \{a,c\}\{a,c\} \match ca$,
even though $\{a,b\}c \not\match ca$.

%BILL
However, note that the very concept of ``regularity of a string" in some
sense breaks down when we consider the concept of a sliding cover: now the
``cover" need not actually ``match" the area it is covering. In fact, the above
concept even allows for a string to be a cover of an indeterminate string
without being a substring of the latter at all! This motivates the idea of a
\itbf{rooted cover} of length $\kappa$, where every covering substring is
required to match, not the preceding entry in the cover, but rather the prefix
of \s{x} of length $\kappa$.
A rooted cover is defined simply by changing ``suffix'' to ``prefix''
in part (b) of Definition~\ref{defn-sliding}.
The example string (\ref{x'}) has no rooted cover,
but the string $\s{x'} = \{a,b\}c\{a,c\}\{a,c\}ac$
has both a sliding cover and a rooted cover of length 2. Notably, in the
literature, the concept of rooted cover is in fact used as the cover for an
indeterminate string \cite{BRS09}.

%\add[S]{In the following section we will present an algorithm to compute the rooted covers of an indeterminate string given its prefix array. In what follows, the following definition and observation will be handy.}
%\begin{definition}
%\label{defn-active}
%A position $i$ in $\s{x}[1..n]$ is said to be \itbf{$c$-active}, $c \le i$,
%if and only if
%\begin{itemize}
%\item[(a)]
%$c = i$; or
%\item[(b)]
%$\s{x}[1..i]$ has a $c$-cover.
%\end{itemize}
%\end{definition}
%\begin{lemma}
%\label{lemma:active}
%If positions $i,i\+ 1,\ldots,i\+ c\- 1$ are not active,
%then no position $j \ge i\+ c$ is active.
%\end{lemma}
% \begin{definition}
% \label{defn-active}
% A position $i$ in $\s{x}[1..n]$ is said to be \itbf{$c$-active}, $c \le i$,
% if and only if
% \begin{itemize}
% \item[(a)]
% $c = i$; or
% \item[(b)]
% $\s{x}[1..i]$ has a $c$-cover.
% \end{itemize}
% \end{definition}
%
% \begin{lemm}
% \label{lemm-active}
% If positions $i,i\+ 1,\ldots,i\+ c\- 1$ are not active,
% then no position $j \ge i\+ c$ is active.
% \end{lemm}

%\section{Computing Rooted Covers}
\subsection{Computing Rooted Covers}
In this section we describe Algorithm $\PCInd$ (Fig. \ref{algo-position-array})
%BILL
to compute  the set of rooted covers $\Gamma$ of a given indeterminate string
$\s{x} \in \Sigma^n$ directly from its prefix table. As will be shown
below, the algorithm runs in linear time on average and $O(n^2)$ time in the
worst case.

Algorithm $\PCInd$ maintains a list $\mathcal L$ to store the candidate rooted
covers. The algorithm also maintains an auxiliary  push-down store $\mathcal
D$, which stores the list of dead covers at each iteration $i \in [2..n]$.  The
push-down store $\mathcal D$ will be used for marking the dead covers so as to
delete them at the end of each iteration. Lastly, in order to determine whether
or not the cover of length $v$ is live at position $i$, the algorithm maintains
an array $maxlive[1..n]$ the same as in Algorithm $\PCR$.

\begin{figure}[t!]
 %\begin{algorithm} [t!]
\begin{algorithmic}
\Procedure{\bf{$\PCInd$}}{$\pi, \Gamma$}
\State $\Gamma \la \phi$; \ $\mathcal L \la \phi$; \ $maxlive[1..n] \la
0^n$ \State $max \la \max(\pi[2..n])$
\LineComment{fill the list $\mathcal L$ with the candidate covers from
$\{1,2,\ldots,max\}$}
\For{$i \la 1 \ \bto max$}
\LineComment{consider only border values}
\If{$\pi[i]+i-1 = |\s{s}|$}
\State $\mathcal L \xleftarrow{+} i$
\EndIf
\EndFor
\For{$i \la 2 \ \bto n$}
\LineComment{$\mathcal D$ stores list of dead covers at position $i$}
\State $\mathcal D \la \phi$
%\State $p \la \pi[i]$; \ $q \la \mathcal L[j]$
\ForAll{$(v \in \mathcal L)$}
\LineComment{skip values of $v > \pi[i]$} 
\If{$(v > \pi[i])$}
\State \bbreak
\EndIf
\State $t \la i + v - 1$
\If{($(maxlive[v] = 0 \ \band t \leq 2 * v)$ \State \bor
$(maxlive[v] \ge t - v)$)}
\LineComment{cover $v$ is still live}
\State $maxlive[v] = t$
\Else
\LineComment{cover $v$ is dead}
\State $maxlive[v] = -1$
\LineComment{mark cover $v$ for deletion}
\State $\push(\mathcal D) \la v$
\EndIf
%\State $j \la j-1$
%\State $q \la \mathcal L[j]$
\EndFor
\LineComment{remove the dead covers from $\mathcal L$}
\While{$\top(\mathcal D) \not= \emptyset$}
\State $r \leftarrow \pop(\mathcal D)$
\State $\mathcal L \xleftarrow{-} r$
\EndWhile
\EndFor
\LineComment{report the rooted covers}
\For{$i \la 1 \ \bto n$}
\If{$maxlive[i] = n$}
\State $\Gamma \xleftarrow{+} i$
\EndIf
\EndFor
%\State \Return  $\gamma$
\EndProcedure
\end{algorithmic}
%\end{algorithm}
\caption{Compute all rooted covers of indeterminate string from its prefix
array.}
\label{algo-position-array}
\end{figure}

Exploiting the fact that the rooted cover of an
indeterminate string \s{x} is also a border of it, the algorithm
starts by identifying the set of candidate (rooted) covers as defined below.
 
\begin{definition}
Let $\s{x} \in \Sigma^n$ and let $\pi[1..n]$ be its prefix array. Then the
set of candidate (rooted) covers $\mathcal L$ of the whole string $\s{x}$ is:

\begin{equation} \label{candidates}
\mathcal L \subseteq \pi: \ \text{where} \  \pi[i] + i-1 = n \
\text{for} \ 2 \le i\le n
\end{equation}
\end{definition}

To populate the list of candidate covers, we start by computing the value
%BILL
$max = \max(\pi[2..n])$. Then the algorithm initializes the list $\mathcal
L$ with the filtered entries from the set $\{1,2,...,max\}$, such that $\mathcal
L$ will only store the values that satisfies $y[i] + i-1 = n$ for $i \in
[2..n]$.
%from the set ${1,2,3,..max}$ at the beginning of the algorithm I added a
% loop (of linear cost) to then

During the execution of the main \bfor loop, at each position $i \in [2..n]$.
%BILL
The algorithm tests, for each candidate cover $v$ in list $\mathcal L$, whether
or not $v$ is active.
Based on the result of this test the algorithm appropriately updates the
corresponding entry in the $maxlive$ array and marks the dead covers at position
$i$, by storing those in $\mathcal D$ which will be
deleted at the end of each iteration using a \bwhile loop.

After computing the array $maxlive$ (at the end of the main \bfor loop), we can
easily identify and report the set of rooted covers of the whole string $\s{x}$
simply by finding all the entries in the array $maxlive$ that have the value $n$
(i.e., all entries of the list of candidate covers that are still active).

%\enote[S]{I don't fully understand how you can only loop through $\mathcal
% D$ to delete items in $\mathcal L$. Could you explain a bit more? Can you
% show me the actual code so that I can understand?} \change{Final brief 
% note,}{
A final note regarding the use of the push-down store $\mathcal D$ is in
order. The standard approach, when the programming
%BILL
language in use allows it, is to delete some elements from a list while iterating through it.
%BILL
This can be done either: (1) by iterating backwards through the list and then
deleting within the \bfor loop, or (2) by identifying all items that need
to be deleted and marking them with a flag (in the first iteration), then (in
the second iteration) removing all those items which are flagged for
deletion. However, in  both cases (1) and (2), the algorithm must loop
through all the items in the list $\mathcal L$ after each iteration.
Alternatively, keeping track of the items to remove in another list (e.g.,
in $\mathcal D$) and then, after all items have been processed, enumerating the
remove list ($\mathcal D$) and removing each item from the list of candidate
covers ($\mathcal L$) requires only looping through $\mathcal
D$. 

\subsection{Analysis}
Finding the value $max$ in $\pi[2..n]$ can be done with a simple linear
scan of the array $\pi$. Computing the list $\mathcal L$ of candidate covers can be
done in $O(n)$ time. The main \bfor loop will be executed exactly $n$ times.

Within the loop the checking of the condition whether a cover is active or not
can be done in constant
%BILL
time for a particular value and hence the total testing of $live$ or $dead$ for
all candidate covers requires time proportional to $|\mathcal L|$, which is
$O(n)$ in the worst case. Note that the list $\mathcal L$ tends to get smaller
and smaller as the iteration continues, because we keep removing dead covers
from it after each iteration. However, the complexity remains $O(n)$ in the
worst case (e.g., $\s{x}= a^n$).

Turning our attention to the \bwhile loop at the end of each iteration of
the main \bfor loop, the processing of $\mathcal D$ to remove the dead
covers also requires time proportional to
$\mathcal D$, thus $O(n)$ in the worst case since the total number of
covers is bounded by $n$. We conclude that the worst-case time requirement
for the main \bfor loop is $O(n^2)$. The
final \bfor loop to report the list of rooted covers requires
time proportional to $|maxlive|$ which is $O(n)$.
The algorithm requires linear extra space to store the lists
$maxlive$, $\mathcal L$ and $\mathcal D$. So we have the following result:

\begin{theorem}
Given the prefix table $\pi$ of an indeterminate string $\s{x} =
\s{x}[1..n]$, Algorithm $\PCInd$ correctly computes the set of rooted covers
of the whole string of \s{x} in $O(n^2)$ time and  linear space.
\end{theorem} 

Finally, Bari et. al. \cite{BRS09} proved that the expected number
of borders of an indeterminate string is bounded by a constant.
Since, in the beginning of Algorithm $\PCInd$ we include only the borders in
$\mathcal L$, this means that the size of the list $\mathcal L$ and also
$\mathcal D$ is bounded by a constant. Therefore, based on the analysis
presented above we can conclude that Algorithm $\PCInd$ runs in linear time
on average.

\subsection{An Illustrative Example}
Suppose $\pi= \{12,3,2,1,1,7,6,1,0,3,0,1\}$.
We have $max =7$. The simulation of the algorithm is shown in Fig.
\ref{PCInd-ex}. The algorithm initializes the set $\mathcal L$ with the set of 
candidate covers. Hence, we have $\mathcal L= \{1,3,6,7\}$. At iteration 
$i=6$, we can see that cover $3$ becomes non-active, so the value $maxlive[3]$
is set to $-1$ and the cover $3$ is removed from the set of candidate covers.
Similarly, at iteration $i=10$, the cover $1$ becomes
non-active, so the value $maxlive[1]$ is set to $-1$ and the cover $1$ is
removed from the set of candidate covers. After computing the array $maxlive$,
the list of rooted covers can be identified as all the positions $i$ in
$maxlive$ where $maxlive[i] = n$. So the covers are $6$ and $7$ since
$maxlive[6]= 12$ and $maxlive[7]=12$. We have $\Gamma = \{6, 7\}$.

\begin{figure}[htbp]
\centering 
\begin{tabular}{r|l|l}
	$\s{i}$ ~ & ~ $\s{maxlive}$ ~ & ~ $\s{\mathcal L}$ \\
	\hline
	2 ~ & ~ $\{2,0,4,0,0,0,0,0,0,0,0,0\}$ ~ & ~  $\{1,3,6,7\}$ \\
	3 ~ & ~ $\{3,0,4,0,0,0,0,0,0,0,0,0\}$~ & ~ $\{1,3,6,7\}$ \\
	4 ~ & ~ $\{4,0,4,0,0,0,0,0,0,0,0,0\}$~ & ~ $\{1,3,6,7\}$ \\
	5 ~ & ~ $\{5,0,4,0,0,0,0,0,0,0,0,0\}$~ & ~ $\{1,3,6,7\}$ \\
	6 ~ & ~ $\{6,0,-1,0,0,11,12,0,0,0,0,0\}$~ & ~ $\{1,6,7\}$ \\
	7 ~ & ~ $\{7,0,-1,0,0,12,12,0,0,0,0,0\}$~ & ~ $\{1,6,7\}$ \\
	8 ~ & ~ $\{8,0,-1,0,0,12,12,0,0,0,0,0\}$~ & ~ $\{1,6,7\}$ \\
	9 ~ & ~ $\{8,0,-1,0,0,12,12,0,0,0,0,0\}$~ & ~ $\{1,6,7\}$ \\
	10 ~ & ~ $\{-1,0,-1,0,0,12,12,0,0,0,0,0\}$~ & ~ $\{6,7\}$ \\
	11 ~ & ~ $\{-1,0,-1,0,0,12,12,0,0,0,0,0\}$~ & ~ $\{6,7\}$ \\
	12 ~ & ~ $\{-1,0,-1,0,0,12,12,0,0,0,0,0\}$~ & ~ $\{6,7\}$
\end{tabular}
\caption{The running values of Algorithm $\PCInd$ for a given string
with prefix array $\pi= \{12,3,2,1,1,7,6,1,0,3,0,1\}$}
\label{PCInd-ex} 
\end{figure}

\subsection{The experiment}
To get an idea of how the algorithm behaves in practice, we have
implemented Algorithm $\PCInd$ and conducted a simple experimental study. The
experiments have been carried out on a Windows Server
2008 R2 64-bit Operating System, with Intel(R) Core(TM) i7 2600 processor @
3.40GHz having an installed memory (RAM) of 8.00 GB. The algorithm have
been implemented in $C\#$ language using Visual Studio 2010.

\begin{figure}[h!]  
  \centering
  \includegraphics*[scale=0.6]{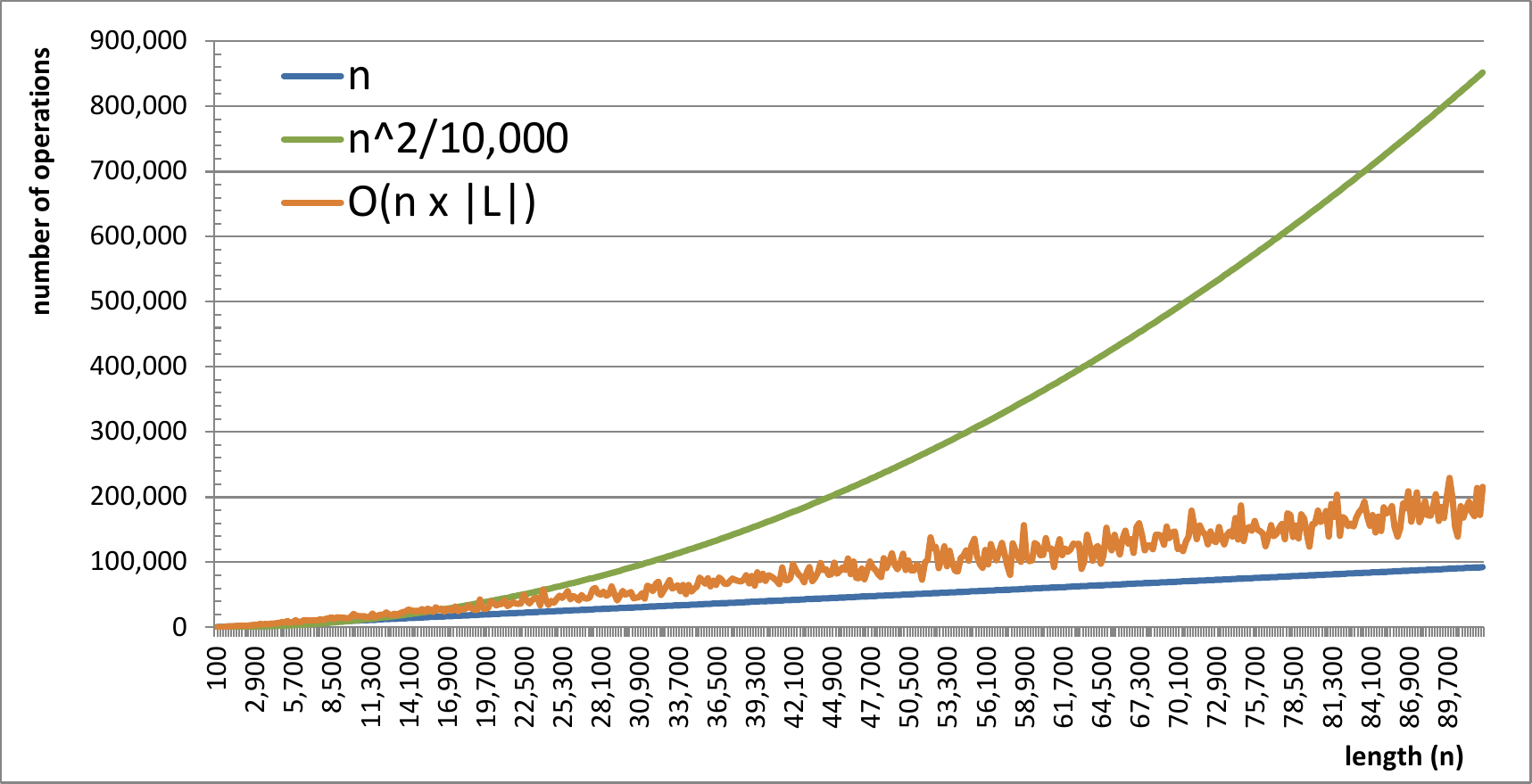}
  \caption{The average running time of the Algorithm $\PCInd$.}
  \label{exp} 
\end{figure} 

We have run Algorithm $\PCInd$ on a set of $100$ randomly  generated prefix
arrays for each length $n \in \{100,200,\ldots,100,000\}$ (averaged over $100$
runs for each length) and counted the average number of executions of the inner
loop of the algorithm. The resulting graph (Fig. \ref{exp}) shows the average
complexity of Algorithm $\PCInd$ fluctuating around $n$. Note that the values
$n^2$ in the graph are scaled down by $10,000$ (i.e., the curves are showing
$n^2/10,000$) to have a better view of the curves. The results show that the
run time of the algorithm is close to linear confirming the average case time
complexity of $O(n)$.

\section{Future Directions}\label{sect-future}
There are several data structures related to the cover array
whose computation may now be contemplated in the context of
indeterminate strings.
For example, a recent paper \cite{FIKPPST13} introduces new forms
of ``enhanced'' cover array that are efficiently computed using the border array;
using the cover array instead would open the way for computation of variants
of these structures also for indeterminate strings.
Similarly, another recent paper \cite{CCIKPRRSW11} proposes efficient
algorithms for the computation of ``seed'' arrays
(a \itbf{seed} of a string \s{x} is a cover of some superstring of \s{x}) --- these
algorithms also may be similarly extended.

\def\AJC{Australasian J.\ Combinatorics\ }
\def\AWOCA{Australasian Workshop on Combinatorial Algs.}
\def\CPM{Annual Symp.\ Combinatorial Pattern Matching}
\def\COCOON{Annual International Computing \& Combinatorics Conference}
\def\FOCS{IEEE Symp.\ Found.\ Computer Science}
\def\AESA{Annual European Symp.\ on Algs.}
\def\LATA{Internat.\ Conf.\ on Language \& Automata Theory \& Applications}
\def\IWOCA{Internat.\ Workshop on Combinatorial Algs.}
\def\AWOCA{Australasian Workshop on Combinatorial Algs.}
\def\STACS{Internat.\ Symp.\ Theoretical Aspects of Computer Science}
\def\ICALP{Internat.\ Colloq.\ Automata, Languages \& Programming}
\def\IJFCS{Internat.\ J.\ Foundations of Computer Science\ }
\def\ISAAC{Internat.\ Symp.\ Algs.\ \& Computation}
\def\SPIRE{String Processing \& Inform.\ Retrieval Symp.}
\def\SWAT{Scandinavian Workshop on Alg.\ Theory}
\def\PSC{Prague Stringology Conf.}
\def\WALCOM{International Workshop on Algorithms \& Computation}
\def\ALG{Algorithmica\ }
\def\CSUR{ACM Computing Surveys\ }
\def\FI{Fundamenta Informaticae\ }
\def\IPL{Inform.\ Process.\ Lett.\ }
\def\IS{Inform.\ Sciences\ }
\def\JACM{J.\ Assoc.\ Comput.\ Mach.\ }
\def\CACM{Commun.\ Assoc.\ Comput.\ Mach.\ }
\def\MCS{Math.\ in Computer Science\ }
\def\NJC{Nordic J.\ Comput.\ }
\def\SICOMP{SIAM J.\ Computing\ }
\def\SIDMA{SIAM J.\ Discrete Math.\ }
\def\JCB{J.\ Computational Biology\ } 
\def\JA{J.\ Algorithms\ }
\def\JCMCC{J.\ Combinatorial Maths.\ \& Combinatorial Comput.\ }
\def\JDA{J.\ Discrete Algorithms\ }
\def\JALC{J.\ Automata, Languages \& Combinatorics\ }
\def\SODA{ACM-SIAM Symp.\ Discrete Algs.\ }
\def\SPE{Software, Practice \& Experience\ }
\def\TCJ{The Computer Journal\ }
\def\TCS{Theoret.\ Comput.\ Sci.\ }


\begin{thebibliography}{99}
\bibitem{A87} Karl Abrahamson,
{\bf Generalized string matching}, {\it \SICOMP 16--6} (1987)
1039--1051.
\bibitem{AHU74} Alfred V.\ Aho, John E.\ Hopcroft \& Jeffey D.\ Ullman,
{\it The Design \& Analysis of Computer Algorithms},
Addison-Wesley (1974).

\bibitem{ARS14} Ali Alatabbi, M.\ Sohel Rahman, \& W.\ F.\ Smyth,
{\bf Inferring an indeterminate string from a prefix graph}, {\it \JDA} (2014),
doi:10.1016/j.jda.2014.12.006.

\bibitem{AE90} Alberto Apostolico \& Andrzej Ehrenfeucht,
{\it Efficient Detection of Quasi-periodicities in Strings}, Tech.\ Report
No.\ 90.5, The Leonardo Fibonacci Institute, Trento, Italy (1990).

\bibitem{AFI91} Alberto Apostolico, Martin Farach \& Costas S.\ Iliopoulos,
Optimal superprimitivity testing for strings, {\it Inform.\ Process.\ Lett.\ 39-1}
(1991) 17-20.

\bibitem{BRS09} Md. Faizul Bari, Mohammad Sohel Rahman \& Rifat Shahriyar, 
{\bf Finding All Covers of an Indeterminate String in O(n) Time on Average}, {\it Stringology} (2009) 263--271.

\bibitem{BKS13} Widmer Bland, Gregory Kucherov \& W.\ F.\ Smyth,
{\bf Prefix table construction \& conversion},
{\it Proc.\ 24th IWOCA},
Springer Lecture Notes in Computer Science LNCS 8288 (2013) 41--53.

\bibitem{BS07} Francine Blanchet-Sadri,
{\it Algorithmic Combinatorics on Partial Words},
Chapman \& Hall/CRC (2008) 385 pp.

\bibitem{B92} D.\ Breslauer,
An on-line string superprimitivity test, {\it Inform.\ Process.\ Lett.\ 44-6} (1992) 345-347.

\bibitem{CRSW13} Manolis Christodoulakis, P,\ J.\ Ryan, W.\ F.\ Smyth \& Shu Wang,.
{\bf Indeterminate strings, prefix arrays \& undirected graphs},
{\it CoRR abs/1406.3289} (2014).

\bibitem{CCIKPRRSW11} Michalis Christou, Maxime Crochemore, Costas S.\ Iliopoulos, Marcin Kubica, Solon P.\ Pissis, Jakub Radoszewski, Wojciech Rytter, Bartosz Szreder \& Tomasz Walen,
{\bf Efficient seeds computation revisited},
{\it Proc.\ 22nd \CPM}, Raffaele Giancarlo \& Giovanni Manzini (eds.),
Lecture Notes in Computer Science, LNCS 6661, Springer-Verlag (2011) 350--363.

\bibitem{FP74} Michael J.\ Fischer \& Michael S.\ Paterson,
{\bf String-matching and other products},
{\it Complexity of Computation, Proc.\ SIAM-AMS 7} (1974) 113-125.

\bibitem{FIKPPST13} Tom\'{a}\u{s} Flouri, C.\ S.\ Iliopoulos, Tomasz Kociumaka,
Solon P.\ Pissis, Simon J.\ Puglisi, W.\ F.\ Smyth \& Wojciech Tyczy\'{n}ski,
{\bf Enhanced string covering},
{\it \TCS 506} (2013) 102--114.

\bibitem{HS03} Jan Holub \& W.\ F.\ Smyth,
{\bf Algorithms on indeterminate strings},
{\it Proc.\ 14th \AWOCA} (2003) 36--45.

\bibitem{HSW08} Jan Holub, W.\ F.\ Smyth \& Shu Wang,
{\bf Fast pattern-matching on indeterminate strings},
{\it \JDA 6--1} (2008) 37--50.
%\bibitem{IP94} Costas S.\ Iliopoulos \&
%Kunsoo Park, An optimal $O(n\log n)$-time algorithm for parallel superprimitivity
%testing, {\it J.\ Korea Information Sci.\ Soc.\ 21-8} (1994) 1400-1404.

\bibitem{LS02} Yin Li \& W.\ F.\ Smyth,
{\bf Computing the Cover Array in Linear Time},
{\it Algorithmica 32--1} (2002) 95--106.

\bibitem{MS94} Dennis
Moore \& W.\ F.\ Smyth, An optimal algorithm to compute all the
covers of a string, {\it Inform.\ Process.\ Lett.\ 50} (1994) 239-246.

\bibitem{MS95} Dennis
Moore \& W.\ F.\ Smyth, Correction to: An optimal algorithm to
compute all the covers of a string, {\it Inform.\ Process.\ Lett.\ 54} (1995) 101-103.

\bibitem{NRR12}	Sumaiya Nazeen, M. Sohel Rahman \& Rezwana Reaz, {\bf Indeterminate string inference algorithms}, 
{\it \JDA 10} (2012) 23--34.
\bibitem{S03} Bill Smyth,
{\it Computing Patterns in Strings},
Pearson Addison-Wesley (2003) 423 pp.

\bibitem{SW08} W.\ F.\ Smyth \& Shu Wang,
{\bf New perspectives on the prefix array}, 
{\em Proc. 15th \SPIRE},
Springer Lecture Notes in Computer Science LNCS 5280 (2008) 133--143.

\bibitem{SW09a} W.\ F.\ Smyth \& Shu Wang,
{\bf A new approach to the periodicity lemma on strings with holes},
{\it \TCS 410--43} (2009) 4295--4302.

\bibitem{SW09} W.\ F.\ Smyth \& Shu Wang,
{\bf An adaptive hybrid pattern-matching algorithm on indeterminate strings},
{\it \IJFCS 20--6} (2009) 985--1004.

\bibitem{VM05} M.\ Vor\'{a}\v{c}ek \& B.\ Melichar,
{\bf Searching for regularities in generalized strings using finite automata},
{\it Proc.\ Internat.\ Conf.\ on Numerical Analysis \& Applied Maths.} Wiley-VCH (2005).
\end{thebibliography}
\end{document}